\documentclass[conference]{IEEEtran}
\IEEEoverridecommandlockouts
\usepackage{cite}
\usepackage{amsmath,amssymb,amsfonts}
\usepackage{algorithmic}
\usepackage{graphicx}
\usepackage{textcomp}
\usepackage{xcolor}
\usepackage{balance}
\def\BibTeX{{\rm B\kern-.05em{\sc i\kern-.025em b}\kern-.08em
    T\kern-.1667em\lower.7ex\hbox{E}\kern-.125emX}}

\usepackage[pscoord]{eso-pic}
\newcommand{\placetextbox}[3]{
\setbox0=\hbox{#3}
\AddToShipoutPictureFG{ \put(\LenToUnit{#1\paperwidth},\LenToUnit{#2\paperheight}){\vtop{{\null}\makebox[0pt][c]{#3}}}
}
}
\placetextbox{.5}{0.060}{\footnotesize { \copyright Authors. Personal use of this material is permitted. Permission must be obtained for all other uses, in any current or future media } }
\placetextbox{.5}{0.050}{\footnotesize {including reprinting/republishing this material for advertising or promotional purposes, creating new  collective works, }}
\placetextbox{.5}{0.040}{\footnotesize {for resale or redistribution to servers or lists, or reuse of any copyrighted component of this work in other works.}}

\begin{document}

\title{Large Language Models in the IoT Ecosystem - A Survey on Security Challenges and Applications\\
}

\author{
\IEEEauthorblockN{
Kushal Khatiwada,
Jayden Hopper,
Joseph Cheatham,
Ayan Joshi, Sabur Baidya
}
\IEEEauthorblockA{
\normalsize{Department of Computer Science and Engineering, University of Louisville, KY, USA}\\
\normalsize{e-mail: \{kushal.khatiwada, jayden.hopper, joseph.cheatham, ayan.joshi, sabur.baidya\}@louisville.edu}
}
}

\maketitle

\begin{abstract}
The Internet of Things (IoT) and Large Language Models (LLMs) have been two major emerging players in the information technology era. Although there has been significant coverage of their individual capabilities, our literature survey sheds some light on the integration and interaction of LLMs and IoT devices - a mutualistic relationship in which both parties leverage the capabilities of the other. LLMs like OpenAI's ChatGPT, Anthropic's Claude, Google's Gemini/BERT, any many more, all demonstrate powerful capabilities in natural language understanding and generation, enabling more intuitive and context-aware interactions across diverse IoT applications such as smart cities, healthcare systems, industrial automation, and smart home environments. Despite these opportunities, integrating these resource-intensive LLMs into IoT devices that lack the state-of-the-art computational power is a challenging task. The security of these edge devices is another major concern as they can easily act as a backdoor to private networks if the LLM integration is sloppy and unsecured. This literature survey systematically explores the current state-of-the-art in applying LLMs within IoT, emphasizing their applications in various domains/sectors of society, the significant role they play in enhancing IoT security through anomaly detection and threat mitigation, and strategies for effective deployment using edge computing frameworks. Finally, this survey highlights existing challenges, identifies future research directions, and underscores the need for cross-disciplinary collaboration to fully realize the transformative potential of integrating LLMs and IoT.
\end{abstract}


\section{Introduction}
A Language Model (LM) is a computational model that is designed to interpret and replicate human language. The primary function of an LM is to identify word patterns and predict the future sequence of input words. A Large Language Model (LLM) is a type of LM characterized by larger parameter sizes and more advanced learning abilities [26]. By leveraging transformer architectures and extensive parameter spaces—often exceeding hundreds of billions of parameters—these models can comprehend context, reason through complex problems, and generate contextually appropriate responses. Some prominent examples of LLM's include OpenAI's GPT series, Anthropic's Claude, and Google's Gemini.

N-Gram models are one of the most common kinds of LM. N-Gram models use context to predict the probability of the next word in a sequence . LLM's have a module called a transformer that allow for the efficient handling of sequential data. In context learning allows the model to produce new text from preexisting text or user given prompts, which enhances the quality of responses. Models are often trained using a technique known as reinforcement learning.  Reinforcement learning utilizes rewards and penalties, similar to human behavioral development, to improve response quality over time [16].

There are a wide variety of applications of LLM in Internet of Things Applications. Researchers have envisioned that combining LLMs with IoT can increase the functionality, intelligence, and autonomy of IoT networks [3]. One such use is in communications and networking applications. LLM's can be used to utilize natural language to simplify and optimize input tasks [1]. Another application of LLM's is smart cities. Smart cities utilize technology such as IoT to improve urban life in a variety of areas, including safety, accessibility, and sustainability. AI systems powered by LLM can enhance all of the previous smart city applications and more [4].

\section{Methodology}
This literature survey aims to provide a comprehensive overview of the integration of Large Language Models (LLMs) within the Internet of Things (IoT) ecosystem, focusing on applications, security aspects, and inherent challenges. To achieve this, a systematic literature search was conducted using prominent academic databases, including IEEE Xplore, ACM Digital Library, Google Scholar, and arXiv.

The selection process prioritized peer-reviewed journal articles, conference papers, and significant preprints published primarily between 2020 and mid-2024 to capture the most recent advancements in this rapidly evolving field. Inclusion criteria focused on works explicitly discussing the synergy, application, or challenges of using LLMs in conjunction with IoT systems. The final selection comprises 30 key papers (as listed in the bibliography) that represent a diverse range of applications, theoretical frameworks, practical implementations, and identified challenges within the LLM-IoT landscape. This methodology ensures a relevant and contemporary synthesis of the current state-of-the-art, although the rapid pace of research means some very recent or niche works might not be included.

\section{Background and Related Work}

\subsection{Evolution of Large Language Models}
Early language models were based on statistical and neural approaches that operated on limited context windows. For example, n-gram models predicted words using the preceding few words in a sequence, but they struggled with long-range dependencies. Recurrent Neural Networks and LSTMs later introduced learned representations of language, yet they too faced challenges in capturing very long contexts and could not be scaled easily in parallel. The transformer architecture revolutionized natural language processing by enabling efficient handling of sequential data through self-attention mechanisms [27]. This architecture forms the backbone of current state-of-the-art LLMs, allowing them to capture long-range dependencies and contextual relationships in text.

\subsection{Evolution of IoT Devices and Ecosystem}

The Internet of Things has evolved from early networked embedded systems into a vast interconnected ecosystem of smart devices. Over the past decades, IoT deployments have expanded from simple sensor nodes and RFID tags to include billions of devices across domains such as transportation, healthcare, industry, and smart homes [3]. By 2023 the number of connected IoT devices worldwide exceeded 16 billion, and this figure is projected to reach 30 billion by 2030 [25]. This growth has been fueled by continual advances in connectivity and computing infrastructure. Modern wireless standards (e.g., 5G and low-power wide-area networks) enable reliable, low-latency communication for massive numbers of devices, while cloud and fog/edge computing frameworks allow data from IoT sensors to be processed closer to the source for real-time responsiveness [4]. Together, these developments have dramatically increased the scale, speed, and capabilities of IoT networks. 

The rapid expansion of IoT has brought corresponding challenges in data management and security. IoT deployments now generate enormous volumes of sensor data, driving the need for scalable data processing and analytics pipelines to extract timely insights. At the same time, security and privacy have become critical concerns, as many IoT devices have simple designs and limited computational resources that make robust security hard to guarantee [22]. The surge in IoT adoption has been accompanied by a rise in cyber threats targeting IoT systems, from device-level attacks to large-scale network intrusions [25]. In response, considerable research has focused on strengthening IoT security through lightweight encryption/authentication protocols and intelligent threat detection. For instance, machine learning-based anomaly detection and intrusion detection systems are increasingly used to identify suspicious behavior in IoT data streams. Overall, the IoT ecosystem has matured with better connectivity and data handling capabilities, but ensuring security and reliability remains an ongoing priority as devices continue to proliferate.

\subsection{Convergence of LLMs and IoT}
With the parallel advancements of LLMs and IoT, researchers have begun exploring their convergence to create smarter, more autonomous systems. The core idea is that LLMs can serve as intelligent interpreters or coordinators for IoT environments, leveraging their language understanding and reasoning abilities to enhance IoT functionalities [3]. By integrating LLMs with IoT, we can enable more natural human–IoT interactions (for example, controlling smart devices through conversational queries) and more adaptive machine–machine communication. Indeed, the research community anticipates that embedding LLM intelligence within IoT networks will improve automation, decision-making, and context-awareness in applications ranging from smart homes to industrial sensor networks [3]. The LLM acts as a high-level brain that can synthesize information from many distributed devices and respond with coherent, informed actions or insights. Initial studies have demonstrated that LLMs can interpret IoT data and even generate code or control logic on the fly, effectively bridging high-level semantic reasoning with low-level device signals [6]. 

Several recent works illustrate the integration of LLMs and IoT across different domains. In smart home scenarios, LLM-based agents have been used as home assistants that interpret user requests and manage connected appliances using the model’s commonsense reasoning and vast knowledge [28]. For example, Rivkin et al. present a Smart Home Agent with Grounded Execution that utilizes an LLM to plan and invoke device-specific actions in response to natural language commands, significantly outperforming earlier rule-based approaches [28]. In the industrial IoT domain, LLMs have been applied to analyze sensor data and coordinate complex tasks; Ren et al. propose combining LLM reasoning with reinforcement learning to optimize industrial processes, highlighting the potential of LLMs for decision-making on the factory floor [y. ren]. Another line of research focuses on enabling LLMs to directly understand IoT sensor inputs. IoT-LM [2] by Mo et al. released in 2024, is one such multisensory model that conditions a pre-trained language model on streams of IoT data, achieving strong performance on sensor-based classification and question-answering tasks [2]. Similarly, Zhou et al. introduce the TENT framework, which aligns IoT sensor signals with textual embeddings so that a frozen LLM can perform human activity recognition in a zero-shot manner (i.e., recognizing new activities without dedicated training data) [6]. This approach endows an IoT system with human-like cognitive flexibility, allowing it to “guess” unseen activities by matching sensor patterns to the closest known semantic descriptions [6]. Beyond automation and data analysis, LLMs are also being leveraged to bolster IoT security. For instance, one study used a GPT-based agent to generate mitigation suggestions for vulnerabilities detected in IoT devices, demonstrating how LLMs can assist in real-time threat response [3]. Another work employs LLM-driven analysis of network logs to detect anomalies in IoT traffic, taking advantage of an LLM’s ability to understand contextual clues in device communications [25]. In summary, the convergence of LLMs and IoT is an emerging area of research, with early results showing that LLMs can significantly enhance IoT applications – from smart cities and homes to industrial and security settings – by providing a layer of semantic intelligence on top of ubiquitous connected devices. Each of these efforts contributes to a growing framework for seamlessly combining the strengths of LLMs (in reasoning and language) with the strengths of IoT (in sensing and actuation), setting the stage for more intuitive and autonomous IoT systems.

\section{Applications of LLMs in IoT}
\subsection{Smart Cities}
Urban IoT deployments (traffic sensors, public services, infrastructure monitors, etc.) can benefit from LLM-driven analysis and control. LLMs are regarded as “indispensable sentinels” for ensuring the seamless operation of complex smart city systems alongside other AI technologies [4]. For example, an LLM could analyze city-wide sensor data and citizen feedback to help city managers optimize services. While computationally heavy, this process can be optimized by pre-processing the data and ensuring that bias, sample size, and other metrics are taken into account. Recent work highlights the substantial potential of LLMs in optimizing ICT processes within smart cities, making urban services more responsive and sustainable [4]. By understanding natural language inputs from city administrators or residents, LLMs could facilitate smart city management tasks such as interpreting maintenance requests, analyzing urban data streams, and coordinating responses across city departments. 

A key challenge in smart city adoption is public opinion and sentiment. City governments and urban planners must work to inform citizens of the benefits of smart city infrastructure and convince them to support the implementation of said infrastructure. A key step in aligning city residents with smart city goals is top-down communication. Top-down communication can highlight the policy and economic priorities of a city. In the study conducted by Nicolas, Kim, and Chi, topics were separated into 6 key domains: smart economy, smart people, smart governance, smart mobility, smart environment, and smart living. The official announcements for a set of cities were then compiled via web scraping to organize them into the previously mentioned domains. Across the cities analyzed, smart economy and smart living were the most frequent domains being communicated about [5]. The study also revealed that multi-disciplinary communication (that is, communications that blended elements of different domains) were most common in mature cities like Boston and Seoul [5]. LLM's can be used to improve city governance and communication, as seen in section 3.3 of the paper by Ullah et al., which discusses smart governance [4].

\subsection{Healthcare}
In smart healthcare and medical IoT settings, LLMs act as intelligent assistants for both patients and providers. They have shown success in answering medical queries and supporting clinical decision-making in all specialties [4]. For example, an LLM-based system can ingest data from wearable health monitors or hospital IoT devices and provide insights or recommendations in plain language. One scenario demonstrated that an LLM could keep track of patients’ treatment schedules via IoT sensors and send timely medication reminders [3]– effectively serving as a smart healthcare assistant. De Vito [21] evaluated healthcare software that integrates IoT data with LLM technology, reflecting growing interest in clinical applications. For instance, an LLM could analyze a diabetic patient’s glucose monitor readings alongside diet and exercise logs (as text) and provide recommendations or warnings in natural language. Another example is using LLMs to interpret complex sensor data like ECG signals or radiology reports by translating them into textual summaries or actionable insights. Prior work with smaller language models (e.g. BERT) showed success in summarizing radiology reports [13]; extending this with LLMs could enable automated medical reporting from IoT sensors.

The “Penetrative AI” concept by Xu et al. demonstrates that ChatGPT-4 already has embedded world knowledge enabling it to interpret IoT sensor data (like accelerometer or GPS readings) and draw insightful conclusions that usually require expert knowledge [24]. This indicates that future IoT healthcare systems could rely on LLMs to bridge sensor data and medical expertise—e.g., detecting fall incidents from a wearable’s motion data and generating an alert like “Possible fall detected; patient may need help”, with reasoning included. Overall, by translating raw sensor outputs into meaningful assessments, LLMs act as an intelligent layer on top of existing healthcare IoT, making data-driven decisions more transparent and patient-centric.

\subsection{Communications}
Communication network designers are utilizing LLM to develop new communications technology. One such technology is 6G, which is expected to bring better data transfer rates, lower latency, and enhanced device capacity. The development of 6G is being supported through the use of LLM and AI by means of edge intelligence and semantic communication [1]. Edge intelligence refers to the practice of placing LLM / AI on the "edge" of IoT devices like sensors, transmitters, and servers. Semantic communication is an optimization technique in which the meaning of a set of data is transmitted rather than the raw data itself [30]. 

While LLM's show much promise in the field of communications, there are some challenges associated with their implementation. One such challenge is the dynamic nature of network environments. Communication networks often exist in rapidly changing environments, and current LLM's are typically trained for a local problem set, which can cause them to be less effective at dealing with environments that evolve rapidly [1]. Another challenge with implementing LLM technology in communications is that LLM's are typically created with task specific behavior in mind. Communication networks, however, often consist of a large variety of connected devices, each of which have different requirements [1]. 

There are multiple ways in which LLM technology will play a role in the development of 6G technology. First, LLM's can generate large amounts of data that can be used for testing and system design. Due to the fact that this data is generated by the model, it is free of any private or confidential information that would come with traditional data gathering techniques . LLM's can also organize and compile network knowledge for system designers. They can also draw conclusions from existing data that designers might miss upon manual review. LLM technology, especially edge intelligence, can also be used to schedule tasks more efficiently by providing real time adjustments to system design based on task requirements [1]. 

\subsection{Smart Homes}
Smart Homes are leveraging LLM to better smart assistants in interpreting user requests. One such case is with smart home agent with grounded execution. SAGE has the ability to overcome the challenges of lacking specific knowledge on the user and their home through a scheme in which a user request triggers a LLM-controlled sequence of discrete actions. These actions then retrieve information, interact with the user, or modify device states [28]. Therefore, By interpreting user commands and executing actions through a dynamic decision tree, SAGE provides personalized and context-aware responses. Its capabilities extend to state monitoring, user preference management, and API-based device control. Compared to existing LLM-based systems, SAGE achieves a higher success rate in executing diverse smart home tasks, highlighting the transformative role of LLMs in creating intelligent and user-friendly home automation systems.

Through the use of LLMs utilizing all the IoT devices in a home Sasha(smarter smart home assistant) plans to addresses the challenges of processing under-specified user commands like 'make it cozy' by using an LLM-driven reasoning process. This approach overcomes the difficulties systems face when lacking specific knowledge about user intent or device capabilities through a scheme in which a user request triggers an LLM-controlled sequence of reasoning steps. These steps involve clarifying goal achievability, retrieving information, planning actions, and interacting with the user [15]. Therefore, by interpreting under-specified commands through this iterative reasoning process, Sasha generates executable action plans, providing personalized and context-aware responses. Its capabilities extend to reasoning about device settings, creating automation routines, user interaction via feedback, and API-based device control through the execution of its generated plans. Compared to existing LLM applications in smart homes, Sasha's iterative approach aims to reduce wrong outputs and improve relevance.

\subsection{Agriculture}
IoT is reshaping modern agriculture by enabling data-driven decision making and automation. Through the deployment of interconnected sensors, drones, and actuators across farms, real-time data are collected on crucial parameters such as soil moisture and nutrient levels, ambient weather conditions, crop health status and livestock well-being [10]. Key application areas extensively documented include precision crop management, where sensor data informs optimized irrigation scheduling and targeted fertilization; automated pest and disease monitoring systems that leverage imagery and environmental data for early detection and intervention; and intelligent livestock management systems for tracking animal health and behavior. Collectively, these IoT applications facilitate improved operational efficiency, significant optimization of resource use such as water and chemicals, improved yields and quality, and contribute to more sustainable and resilient agricultural practices, addressing critical global food security challenges.

\subsection{Industry}
The use of IoT in industry allows for monitoring, control, and efficiency through interconnected devices and sensors across manufacturing floors, supply chains, and infrastructure. Early applications focused on data collection and basic automation. However, recently there has been a strong push towards integrating artificial intelligence to unlock more capabilities. For example, emerging research exemplified by studies combining IIoT with Large Language Models (LLMs) using intelligence-based Reinforcement Learning (RL) approaches explores applications beyond simple connectivity [18]. This line of inquiry investigates the use of LLMs potentially to interpret complex, unstructured operational data like technician logs or manuals or to facilitate natural language interfaces for diagnostics and control, while employing RL to train intelligent agents for autonomous decision-making in dynamic industrial environments. Specific applications potentially explored within this framework include optimizing complex manufacturing processes through learned control strategies, enabling adaptive predictive maintenance based on multifaceted data streams, facilitating autonomous robotic operations in unstructured settings, and dynamically managing energy consumption or resource allocation in real time, thus showcasing how the synergy between IIoT, LLM, and RL is driving the development of more intelligent, adaptive, and autonomous industrial systems.

\begin{table*}[htbp]
\caption{Overview of LLM Applications in Various IoT Domains}
\begin{center}
\begin{tabular}{|p{1.8cm}|p{3.5cm}|p{4cm}|p{3.5cm}|p{1.5cm}|}
\hline
\textbf{IoT Domain} & \textbf{Specific Application Example} & \textbf{LLM Role / Capability Leveraged} & \textbf{Key Benefits / Challenges Addressed} & \textbf{Ref.} \\
\hline
Smart Cities & Optimizing urban services, analyzing sensor data & Natural language understanding, data synthesis, decision support & Improved service responsiveness, citizen alignment, complex system management & [4], [5] \\ \hline
Healthcare & Patient monitoring, medical query answering, report summarization & Data interpretation (wearables, sensors), plain language generation, knowledge retrieval & Personalized assistance, clinical decision support, automated reporting & [3], [4], [13], [21], [24] \\ \hline
Communications (e.g., 6G) & Network optimization, data generation for testing, semantic communication & Task scheduling, synthetic data generation, understanding communication intent & Enhanced efficiency, reduced latency (semantic comms), aids network design & [1], [30] \\ \hline
Smart Homes & Interpreting user commands, controlling devices, automation & Natural language processing, commonsense reasoning, planning, grounded execution & Context-aware control, personalized automation, handling underspecified requests & [15], [19], [28] \\ \hline
Agriculture & Precision crop management, livestock monitoring & Data analysis (sensors, imagery), decision support & Optimized resource use, improved yields, sustainable practices & [10] \\ \hline
Industry (IIoT) & Process optimization, predictive maintenance, operational data analysis & Reasoning, reinforcement learning integration, complex data interpretation & Enhanced automation, intelligent decision-making, adaptive control & [3], [18] \\ \hline
IoT Security & Anomaly detection, threat mitigation suggestions, device fingerprinting & Contextual understanding (logs), pattern recognition, knowledge-based response generation & Improved threat detection accuracy, real-time assistance, identifying unknown devices & [3], [16], [22], [25] \\ \hline
Cross-Domain & Zero-shot activity recognition from sensor data & Aligning sensor signals with textual embeddings, cognitive flexibility & Recognizing novel activities without specific training & [6] \\ \hline
Cross-Domain & General IoT task reasoning, sensor data understanding & Multisensory data processing, bridging semantic reasoning and device signals & Enhanced real-world interaction, complex task execution & [2], [8], [20] \\
\hline
\multicolumn{5}{l}{Note: This table synthesizes findings; refer to cited papers for specific details and methodologies.}
\end{tabular}
\label{tab:applications}
\end{center}
\end{table*}    

\section{Challenges in LLM-IoT}
\subsection{Latency and Bandwidth Issue}
Integrating Large Language Models (LLMs) with the Internet of Things (IoT) presents significant challenges related to latency and bandwidth. IoT systems often involve numerous devices that generate continuous streams of data that require real-time processing for timely actions, such as in smart homes [28] or industrial environments [18]. LLMs are computationally intensive and large mainly running in the cloud. Transmitting large amounts of sensor data [2] to a remote LLM and waiting for inference results introduces network latency, which can be excessive for time-critical applications such as 6G communications [1] or complex task execution [8]. Furthermore, the sheer volume of data transfer consumes significant network bandwidth. This constraint is relevant in resource-limited IoT deployments or wide-area networks such as those in smart cities [4]. Where approaches like efficient prompting [7] and model orchestration [8] aim to mitigate these issues, the fundamental mismatch between real-time IoT demands and the processing/communication requirements of large models remains a key hurdle, affecting the feasibility of applications requiring LLMs to understand and react to the physical world instantly [24].

\subsection{Data Privacy}
The combination of LLMs and IoT raises concerns about data privacy and regulatory compliance. IoT devices, such as those installed in personal spaces such as smart homes [15] or sensitive sectors such as healthcare care [21], collect large amounts of highly personal data. Feeding these data into LLMs for analysis or control creates significant privacy risks, including potential unauthorized access, data breaches, or misuse of information [16]. Ensuring compliance with data protection regulations (e.g. GDPR, HIPAA) becomes complex, particularly when data traverse networks or involve third-party LLM providers whose data handling practices may not be transparent. There is also the risk of LLMs accidentally leaking sensitive information learned during training or generating responses that reveal private details [23]. Therefore, developing privacy-preserving techniques, such as deploying specialized lightweight models on edge devices [25] or implementing robust data anonymization and encryption protocols, is crucial to build reliable LLM-IoT systems, especially in sectors such as healthcare care [21], cybersecurity [23], and smart cities [5].

\subsection{Cost}
The deployment and operation of LLMs within IoT ecosystems present significant cost management challenges. LLMs require substantial computational resources for training, fine-tuning, and inference, leading to high energy consumption and potentially large cloud computing bills or expensive edge hardware investments [4]. Processing continuous high-volume data streams from potentially thousands or millions of IoT devices further increases these computational demands [2]. Developing and maintaining specialized LLMs tailored for specific IoT applications also requires considerable investment in expertise and development time. Techniques such as efficient prompting or the use of more lightweight models are being explored to reduce operational expenses, but the overall cost of using powerful AI such as LLMs in distributed IoT networks remains a significant barrier, particularly for large-scale deployments in areas like smart agriculture [10] or smart cities [4], which could limit the economic viability and widespread adoption of these integrated systems.

\subsection{Reliability}
While LLMs demonstrate impressive capabilities, they are more susceptible to generating outputs that are plausible but factually incorrect or nonsensical [26]. LLMs may need to interpret sensor data or directly or indirectly manipulate physical actuators, such unreliability poses significant safety risks. An LLM misinterpreting critical sensor readings from industrial machinery or generating incorrect guidance or commands related to healthcare applications could lead to equipment damage, operational failure, or harm to individuals  [21]. Ensuring the LLM outputs are factual and reliable when interacting with the physical world is vital [24]. In order for that a mechanism for validation, uncertainty estimation, fail-safe protocols, and potentially human-in-the-loop verification for safety-critical decisions is needed. Therefore, adding layers of complexity to system design, assessment, and operation.

\begin{table*}[htbp]
\caption{Key Challenges in LLM-IoT Integration and Potential Mitigation Strategies}
\begin{center}
\begin{tabular}{|p{2cm}|p{4.5cm}|p{6.5cm}|p{1.5cm}|}
\hline
\textbf{Challenge} & \textbf{Description} & \textbf{Potential Mitigation Strategies / Research Directions} & \textbf{Ref.} \\
\hline
Latency & Delay in processing sensor data and receiving LLM response, critical for real-time tasks. & Edge computing (processing closer to source), efficient prompting techniques, model quantization/pruning, lightweight LLMs, semantic communication (transmit meaning, not raw data). & [1], [2], [4], [7], [8], [24], [25], [30] \\ \hline
Bandwidth & High volume of sensor data transfer consumes network resources. & Edge computing (local processing/filtering), data compression, semantic communication, selective data transmission. & [1], [2], [4], [8] \\ \hline
Data Privacy & Sensitive personal data collected by IoT devices processed by potentially third-party LLMs. & On-device/Edge LLMs, data anonymization, encryption protocols, federated learning, differential privacy, strict compliance frameworks (GDPR, HIPAA). & [15], [16], [21], [23], [25] \\ \hline
Computational Cost & High resource requirements (energy, compute) for training and inference of large LLMs. & Lightweight models, model optimization (quantization, pruning), hardware acceleration (on edge), efficient prompting, cloud-edge hybrid models. & [2], [4], [10], [25] \\ \hline
Reliability / Accuracy & LLMs can "hallucinate" or generate factually incorrect/nonsensical outputs, risky for physical system control. & Validation mechanisms, uncertainty quantification, human-in-the-loop verification, grounding LLM outputs in physical context, robust testing frameworks. & [21], [24], [26] \\ \hline
Security Vulnerabilities & IoT devices can be attack vectors; LLM integration adds complexity and potential new vulnerabilities. & Lightweight security protocols, LLM-based anomaly detection, secure integration practices, regular vulnerability assessment, robust authentication. & [3], [22], [23], [25], [29] \\ \hline
Scalability & Managing and coordinating LLM interactions with potentially billions of IoT devices. & Efficient orchestration frameworks, hierarchical control structures, standardized APIs, scalable cloud/edge infrastructure. & [8], [25] \\
\hline
\multicolumn{4}{l}{Note: Mitigation strategies are often overlapping and actively researched. Refer to cited papers for specifics.}
\end{tabular}
\label{tab:challenges}
\end{center}
\end{table*}

\section{Discussion}
The integration of Large Language Models (LLMs) with the Internet of Things (IoT) represents a significant paradigm shift, moving beyond simple data collection and actuation towards context-aware, intelligent, and interactive cyber-physical systems. As synthesized in this survey, LLMs act as a "cognitive layer" or "brain" [3], [6], capable of interpreting complex sensor data, understanding natural language commands, reasoning about the physical world [24], and orchestrating actions across diverse IoT devices [8].

Applications span numerous domains (Table \ref{tab:applications}), from enhancing urban living [4], [5] and personalizing healthcare [21] to optimizing industrial processes [18] and securing networks [25]. A key enabler is the LLM's ability to bridge the gap between high-level human intent or semantic goals and low-level device operations [6], [15], [28], often leveraging commonsense reasoning and vast world knowledge. This allows for more intuitive human-IoT interaction and more adaptive autonomous system behavior.

However, realizing this potential is fraught with challenges (Table \ref{tab:challenges}). The inherent resource intensiveness of LLMs clashes with the often-constrained nature of IoT devices and networks, leading to significant concerns regarding latency, bandwidth consumption, and operational cost [1], [4], [25]. Transmitting potentially vast amounts of sensitive sensor data to centralized LLMs raises critical privacy and security issues [16], [21], [23], demanding robust mitigation strategies like edge processing and advanced encryption. Furthermore, the reliability of LLMs remains a concern; the potential for generating incorrect or nonsensical outputs (hallucinations) [26] poses substantial risks in safety-critical IoT applications controlling physical systems [21], [24].

Current research actively seeks to address these limitations. Strategies like deploying lightweight or specialized LLMs at the network edge [25], developing efficient prompting and orchestration techniques [7], [8], and exploring novel communication paradigms like semantic communication [30] are crucial. Security is also paramount, with research focusing both on leveraging LLMs for enhanced IoT security (e.g., anomaly detection [22]) and on securing the LLM-IoT integration itself [23], [29].

The maturity of LLM-IoT integration varies across domains. Smart home applications [15], [28] and certain data analysis tasks appear more advanced, while real-time control in industrial or communication settings [1], [18] faces higher hurdles due to latency and reliability requirements. Overall, the convergence is promising but requires continued interdisciplinary effort spanning AI, networking, security, and domain-specific engineering to overcome the existing barriers and unlock the full transformative potential.    

\section{Future Research Directions}
While significant progress has been made, the seamless and widespread integration of LLMs and IoT necessitates further research across several key areas:

\begin{itemize}
    \item \textbf{Efficient and Edge-Optimized LLMs:} Developing novel LLM architectures specifically designed for resource-constrained edge devices is paramount. This includes techniques beyond simple quantization or pruning, potentially exploring new model types or hardware-software co-design. Research on extremely lightweight models capable of meaningful reasoning on-device is crucial [25].
    \item \textbf{Robustness and Reliability:} Addressing LLM hallucinations and ensuring factual grounding in the physical world is critical for safety-critical applications [24], [26]. Future work should focus on mechanisms for uncertainty quantification, runtime validation of LLM outputs against sensor data, fail-safe protocols, and potentially hybrid systems combining LLMs with traditional control methods.
    \item \textbf{Privacy-Preserving Integration:} Beyond current techniques like encryption and anonymization, research is needed into advanced privacy-preserving machine learning (PPML) techniques like federated learning tailored for LLM-IoT, secure multi-party computation, and verifiable differential privacy guarantees, especially when dealing with sensitive data streams [16], [23].
    \item \textbf{Explainable AI (XAI) for LLM-IoT:} Understanding why an LLM makes a particular decision or generates a specific command for an IoT device is vital for debugging, trust, and accountability. Developing XAI methods suitable for the complex, often black-box nature of LLMs operating in dynamic IoT environments is an important direction.
    \item \textbf{Real-World Benchmarking and Deployment:} Standardized benchmarks and realistic testbeds are needed to evaluate the performance, robustness, and efficiency of different LLM-IoT integration strategies across various applications. More studies focusing on long-term, real-world deployments are required to understand practical challenges [2], [20].
    \item \textbf{Seamless Sensor-Language Grounding:} Improving the ability of LLMs to directly understand and reason about raw, multi-modal sensor data [2] beyond current approaches [6] will enhance their effectiveness in interpreting complex physical world states.
    \item \textbf{Ethical Considerations:} As LLMs become more integrated into personal spaces and critical infrastructure via IoT, research must address the ethical implications, including bias in decision-making, potential for misuse, transparency, and societal impact [4].
    \item \textbf{Scalable Orchestration and Management:} Developing frameworks for efficiently managing potentially millions or billions of IoT devices interacting with LLMs, including discovery, task allocation, and coordination, remains a challenge [8].
\end{itemize}
Addressing these research directions will be key to building truly intelligent, reliable, secure, and beneficial LLM-powered IoT ecosystems.

\section{Conclusion}
The convergence of Large Language Models and the Internet of Things heralds a new era of intelligent, interactive, and autonomous systems. This survey has explored the current state-of-the-art, highlighting the diverse applications where LLMs enhance IoT capabilities – from smart cities and healthcare to industrial automation and smart homes – by providing advanced reasoning, natural language understanding, and context awareness. We have systematically reviewed the significant role LLMs can play in bolstering IoT security through anomaly detection and threat mitigation.

Despite the immense potential, significant challenges persist, primarily concerning latency, bandwidth constraints, computational costs, data privacy, security vulnerabilities, and the inherent reliability of LLMs when interacting with the physical world. Ongoing research is actively developing mitigation strategies, including edge computing frameworks, lightweight models, efficient communication protocols, and robust security measures.

The future of LLM-IoT integration hinges on addressing these challenges through continued innovation in efficient model design, privacy-preserving techniques, reliability mechanisms, and scalable architectures. Cross-disciplinary collaboration is essential to navigate the technical complexities and ethical considerations. Fully realizing the transformative potential of combining the cognitive power of LLMs with the sensing and actuation capabilities of IoT promises to reshape numerous aspects of technology and society, demanding careful and concerted research efforts moving forward.

\balance

\vspace{12pt}

\end{document}